\documentclass{PoS}
\usepackage{graphicx}
\usepackage{epsfig}
\usepackage{color}
\usepackage{longtable}
\usepackage{latexsym}
\usepackage{amsmath}
\usepackage{amssymb}
\usepackage{dsfont}
\usepackage{verbatim}
\usepackage{bm}

\title{Study of the $Z_c^+$ channel in lattice QCD}

\ShortTitle{Study of the $Z_c^+$ channel in lattice QCD}

\author{\speaker{Luka Leskovec}\\
        Jozef Stefan Institute, 1000 Ljubljana, Slovenia\\
        E-mail: \email{luka.leskovec@ijs.si}}

\author{Sasa Prelovsek\\
        Department of Physics, University of Ljubljana and Jozef Stefan Institute, 1000 Ljubljana, Slovenia\\
        E-mail: \email{sasa.prelovsek@ijs.si}}

\author{C.B. Lang\\
        Institut f\"ur Physik, University of Graz, A-8010 Graz, Austria\\
        E-mail: \email{christian.lang@uni-graz.at}}

\author{Daniel Mohler\\
        Fermi National Accelerator Laboratory, Batavia, Illinois 60510-5011, USA\\
        E-mail: \email{dmohler@fnal.gov}}

\abstract{Several charged charmonium-like hadrons called $Z_c$ have been recently discovered by different experiments. In contrast to conventional hadrons these contain at least two valence quarks and antiquarks ($\bar{c}c\bar{d}u$). We perform a lattice QCD simulation of the $I^G(J^{PC})=1^+(1^{+-})$ channel including all relevant two-meson operators under $4.3$ GeV:  $J/\psi\, \pi$, $\psi_{2S}\pi$, $\psi_{1D}\pi$, $D\bar D^*$, $D^*\bar D^*$, $\eta_c\rho$ as well as additional diquark anti-diquark operators. In our $N_f=2$ simulation with pion mass at $266$ MeV we are able to identify all two-meson levels within the energy region of interest. However we find no additional level identifiable as a candidate for $Z_c$.}

\FullConference{The 32nd International Symposium on Lattice Field Theory,\\
		23-28 June, 2014\\
		Columbia University New York, NY}

\begin{document}
\section{Motivation}

Mesons are most often considered as a pair of a quark and anti-quark, but recent experimental data shows that there exist exotic mesons made up from at least two quarks and anti-quarks ($\bar{c}c\bar{d}u$) as reviewed in \cite{Brambilla:2014jmp}. Among such states are some that have negative charge conjugation number and we list those states together with their dominant decay modes and quantum numbers in Table \ref{tab:expstat}.

\begin{table}[h]
\begin{center}
\begin{tabular}{|c|c|c|c|c|} \hline
particle        &  $C$   &   $J^{P}$   &   decay            &   collaboration\\ \hline
$Z^+(4430)$     &  $-1$  &   $1^{+}$   &   $\psi(2S) \pi^+$ &   Belle \cite{Choi:2007wga}, LHCb \cite{Aaij:2014jqa}\\
$Z_c^+(3900)$   &  $-1$  &   $?^{?}$   &   $J/\psi \pi^+$   &   BESIII \cite{Ablikim:2013mio}, Belle
 \cite{Liu:2013dau}, CLEO-c \cite{Xiao:2013iha}\\
$Z_c^+(3885)$   &  $-1$  &   $1^{+}$   &   $(D D^*)^+$      &   BESIII \cite{Ablikim:2013xfr}\\
$Z_c^+(4020)$   &  $-1$  &   $?^{?}$   &   $h_c(1P) \pi^+$  &   BESIII \cite{Ablikim:2013wzq}\\
$Z_c^+(4025)$   &  $-1$  &   $?^{?}$   &   $(D^* D^*)^+$    &   BESIII \cite{Ablikim:2013emm}\\
$Z^+(4200)$     &  $-1$  &   $1^{+}$   &   $J/\psi \pi^+$   &   Belle \cite{Chilikin:2014bkk}\\ \hline
\end{tabular}
\label{tab:expstat}
\end{center}
\caption{Experimental status of the $Z_c$ candidates in the $J^{PC}=1^{+-}$ channel.}
\end{table}
In this study we focus on the $J^{PC}=1^{+-}$ charmonium-like channel, where at least two of the states from Table \ref{tab:expstat} appear in experiment. States with the unknown $J^P$ in Table \ref{tab:expstat} are likely also in the same channel, however their quantum numbers have not been determined yet.\\
There were two previous lattice QCD studies of the $J^{PC}=1^{+-}$ charmonium-like channel. The first study \cite{Prelovsek:2013xba} focused on the region below $4$ GeV, however no additional energy levels appeared in that simulation, nor any noticeable energy shifts. The second lattice simulation \cite{Chen:2014afa} focused on $DD^*$ scattering in the $J^{PC}=1^{+-}$ channel. The authors were able to extract near threshold parameters and claimed to find no candidate.\\
\vspace{-0.6cm}
\section{Lattice setup}
\begin{figure}[htb]
    \centering
    \includegraphics[width=0.45\textwidth]{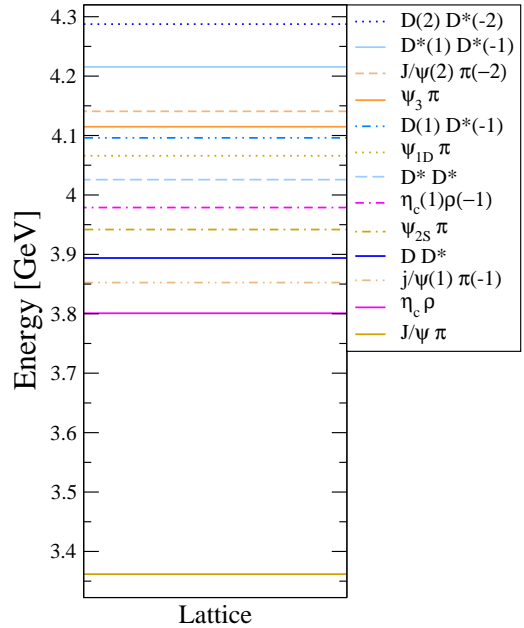}
    \caption{Noninteracting two-meson levels for the $I^G(J^{PC})=1^+(1^{+-})$ channel up to $4.3$ GeV on the $2$ fm lattice. }
    \label{fig:fig1}
\end{figure}

We use an ensemble of gauge configurations with $N_f=2$ Wilson quarks where the pion mass was $266$ MeV and the lattice spacing $a=0.1239(12)$ fm \cite{Hasenfratz:2008fg,Hasenfratz:2008ce}. The full box, $16^3 \times 32$, is about $2$ fm in size, which turns out to be an important advantage, as the amount of two-meson energy levels within an energy region is tractable.\\
The charm quarks are treated with the Fermilab method \cite{ElKhadra:1996mp}, where the discretization effects of the heavy charm quark are suppressed. Energy levels are obtained relative to the spin averaged mass: $E_n - m_{s.a.}$, where $m_{s.a.}=\frac{m_{\eta_c} + 3 m_{J/\psi}}{4}$. In order to show results at the appropriate scales we present our energies as: $E_n - m_{s.a.}^{latt.} + m_{s.a.}^{phys}$.\\
To evaluate the correlator matrix, $C_{ij}=\langle \Omega | {\cal O}_i(t_{snk}) {\cal O}_j^{\dagger}(t_{src})| \Omega \rangle$, we use the distillation method \cite{Peardon:2009gh}, which is an all-to-all method particularly useful on smaller physical volumes. Within this method propagators are translated into perambulators (going from a source field to a sink field). The form of the source particles is encoded within the so-called $\phi$ matrices, which contain all interpolator structure.\\
Operators included in the correlator matrix need to be chosen carefully, as in principle all states with the given quantum numbers appear. This means, that all two-meson states as well as any potential bound states or resonances with $I^G(J^{PC})=1^+(1^{+-})$ are present in the spectrum. In order to be able to recognize any additional states, first all the two-meson energy levels need to be identified. Only after that any additional states can become candidates for $Z_c$. In the simulation, we use $22$ interpolating operators \footnote{$h_c(1)\pi(-1)$ is not included, as the naive implementation couples to the wrong ground state.}, which are listed in Eq. \ref{eq:ops}:
\small
\begin{equation}
\label{eq:ops}
\begin{aligned}
{\cal O}_1^{\psi(0)\pi(0)}&=\bar c \gamma_i c(0)~\bar d\gamma_5 u(0)\,,\\
{\cal O}_2^{\psi(0)\pi(0)}&=\bar c \gamma_i \gamma_t c(0)~\bar d\gamma_5  u(0)\,,\nonumber \\ 
{\cal O}_3^{\psi(0)\pi(0)}&=\bar c \overleftarrow{\nabla}_j\gamma_i \overrightarrow{\nabla}_j c(0)~\bar d\gamma_5 u(0)\,, \nonumber\\
{\cal O}_4^{\psi(0)\pi(0)}&=\bar c \overleftarrow{\nabla}_j\gamma_i\gamma_t \overrightarrow{\nabla}_j c(0)~\bar d\gamma_5 u(0) \,, \nonumber\\
{\cal O}_5^{\psi(0)\pi(0)}&= |\epsilon_{ijk}| |\epsilon_{klm}|~\bar c \gamma_j \overleftarrow{\nabla}_l  \overrightarrow{\nabla}_m c(0)~\bar d\gamma_5 u(0) \,, \nonumber\\
{\cal O}_6^{\psi(0)\pi(0)}&= |\epsilon_{ijk}| |\epsilon_{klm}| ~\bar c \gamma_t \gamma_j  \overleftarrow{\nabla}_l  \overrightarrow{\nabla}_m c(0)~\bar d\gamma_5 u(0)\,, \nonumber\\
{\cal O}_7^{\psi(0)\pi(0)}&=R_{ijk} Q_{klm}~ \bar c  \gamma_j  \overleftarrow{\nabla}_l\overrightarrow{\nabla}_m c~\bar d\gamma_5 u(0)  \,, \nonumber\\
{\cal O}_8^{\psi(0)\pi(0)}&=R_{ijk} Q_{klm}~ \bar c  \gamma_t \gamma_j  \overleftarrow{\nabla}_l\overrightarrow{\nabla}_m c~\bar d\gamma_5 u(0)\,,  \nonumber \\
{\cal O}^{\psi(1)\pi(-1)}_{9}&=\sum_{e_k=\pm e_{x,y,z}}\!\!\!~\bar c \gamma_i c(e_k)~\bar d\gamma_5 u(-e_k)\,, \nonumber\\
{\cal O}^{\psi(2)\pi(-2)}_{10}&=\sum_{|u_k|^2=2}~\bar c \gamma_i c(u_k)~\bar d\gamma_5 u(-u_k)\,, \nonumber\\
{\cal O}^{\eta_c(0)\rho(0)}_{11}&=\bar c \gamma_5 c(0)~\bar d\gamma_i u(0)\,, \nonumber\\
{\cal O}^{\eta_c(1)\rho(-1)}_{12}&=\sum_{e_k=\pm e_{x,y,z}}\bar c \gamma_5 c(e_k)~\bar d\gamma_i u(-e_k)\,, \nonumber\\
\end{aligned}
\begin{aligned}
{\cal O}^{D(0)D^*(0)}_{13}&=\bar c \gamma_5 u(0)~\bar d\gamma_i c(0)   +  \{\gamma_5 \leftrightarrow \gamma_i\}\,,\nonumber  \\
{\cal O}^{D(0)D^*(0)}_{14}&=\bar c \gamma_5 \gamma_t u(0)~\bar d\gamma_i \gamma_t c(0) +   \{\gamma_5 \leftrightarrow \gamma_i\} \,,\nonumber\\
{\cal O}^{D(1)D^*(-1)}_{15}&=\sum_{e_k=\pm e_{x,y,z}}\bar c \gamma_5 u(e_k)~\bar d\gamma_i c(-e_k) +  \{\gamma_5 \leftrightarrow \gamma_i\}   \nonumber\,,\\
{\cal O}^{D(2)D^*(-2)}_{16}&= \sum_{|u_k|^2=2}\bar c \gamma_5 u(u_k)~\bar d\gamma_i c(-u_k) +  \{\gamma_5 \leftrightarrow \gamma_i\}   \nonumber\,,\\
{\cal O}^{D^*(0)D^*(0)}_{17}&=\epsilon_{ijl}~\bar c \gamma_j u(0)~\bar d\gamma_l c(0)\,,   \nonumber\\
{\cal O}^{D^*(1)D^*(-1)}_{18}&=\sum_{e_k=\pm e_{x,y,z}}\epsilon_{ijl}~\bar c \gamma_j u(e_k)~\bar d\gamma_l c(-e_k)   \nonumber\\
{\cal O}^{4q}_{19}&= N_L^3~\epsilon_{abc} \epsilon_{ab'c'}(\bar c_{b}C \gamma_5\bar d_{c}~   c_{b'}\gamma_{i} C u_{c'}\nonumber\\
  &\qquad\qquad\qquad\qquad\quad- \bar c_{b}C \gamma_i\bar d_{c}~  c_{b'}\gamma_{5} C u_{c'})\nonumber \,,\\ 
{\cal O}^{4q}_{20}&= N_L^3~\epsilon_{abc} \epsilon_{ab'c'}(\bar c_{b}C \bar d_{c}~   c_{b'}\gamma_{i} \gamma_5C u_{c'}\nonumber\\
&\qquad\qquad\qquad\qquad\quad- \bar c_{b}C \gamma_i\gamma_5\bar d_{c}~  c_{b'} C u_{c'})\nonumber\,,\\ 
{\cal O}^{4q}_{21}&={\cal O}^{4q}_{19}(N_v\!=\!32)\nonumber\,,\\
{\cal O}^{4q}_{22}&={\cal O}^{4q}_{20}(N_v\!=\!32)\,.\nonumber
\end{aligned}
\end{equation}
\normalsize
The numbers in the parenthesis, i.e. $\bar{c}\gamma_i c (0)$, indicate the momentum projections for each current used in the operators. ${\cal O}_{1-18}^{MM}$ are two-meson operators used to obtain the relevant scattering channels in the energy region up to $4.3$ GeV. Their non-interacting energies $E_1(p) + E_2(-p)$ correspond to the lines drawn in Fig. \ref{fig:fig1}. They represent the two-meson states $J/\psi\, \pi$, $\psi_{2S}\pi$, $\psi_{1D}\pi$, $D\bar D^*$, $D^*\bar D^*$, $\eta_c\rho$ at different momentum projections. There are also four additional operators ${\cal O}_{19-22}^{4q}$, which are the color triplet combinations with "good" and "bad" positive parity diquarks (${\cal O}_{19,21}^{4q}$) and a combination of negative parity diquarks ${\cal O}_{20,22}^{4q}$ \cite{Jaffe:2004ph}. All diquark anti-diquark interpolators are symmetrized to give good charge conjugation.

\begin{figure}[htb]
    \centering
    \hspace{0.6cm}
    \includegraphics[width=0.8\textwidth]{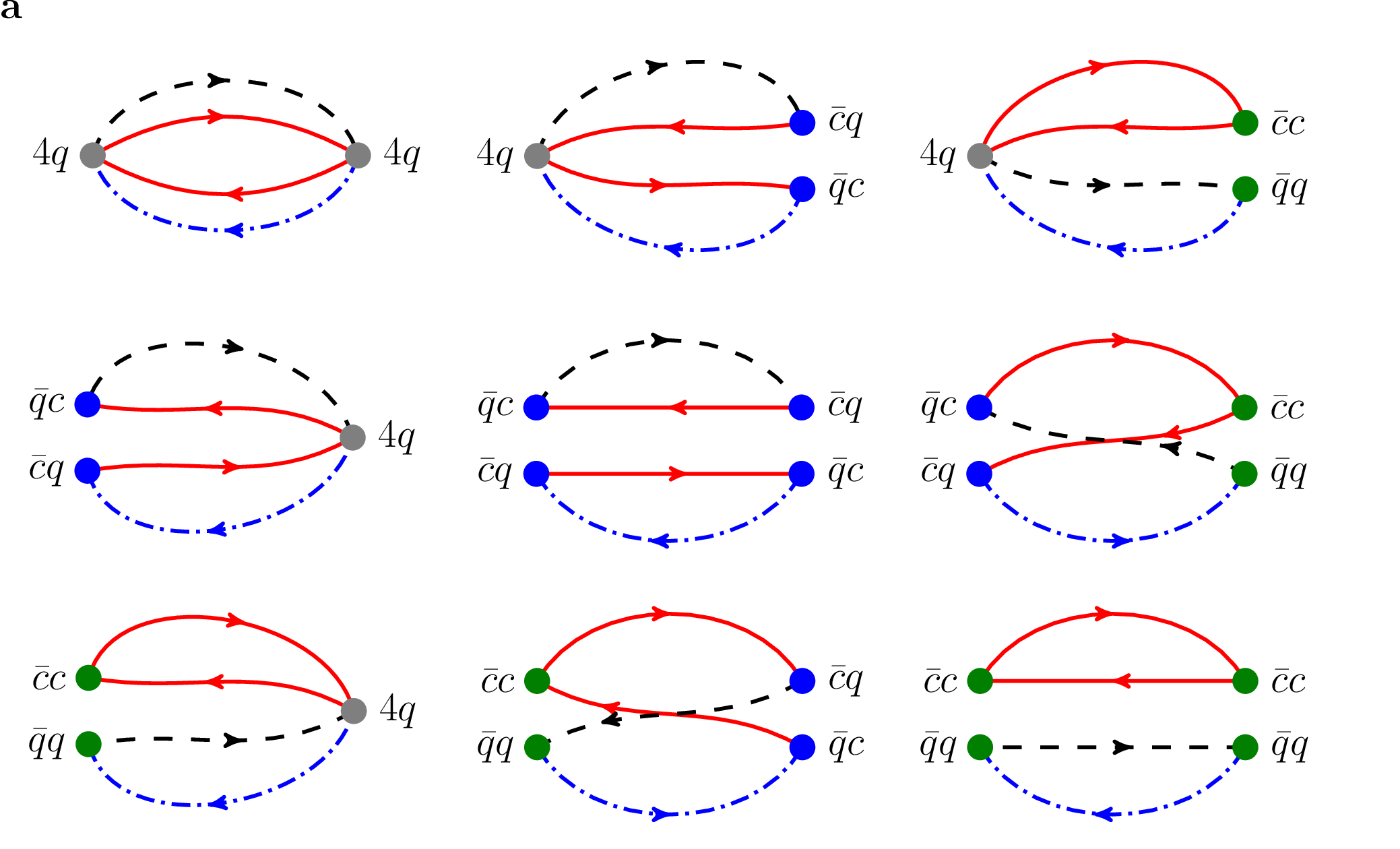}\\
    \includegraphics[width=0.75\textwidth]{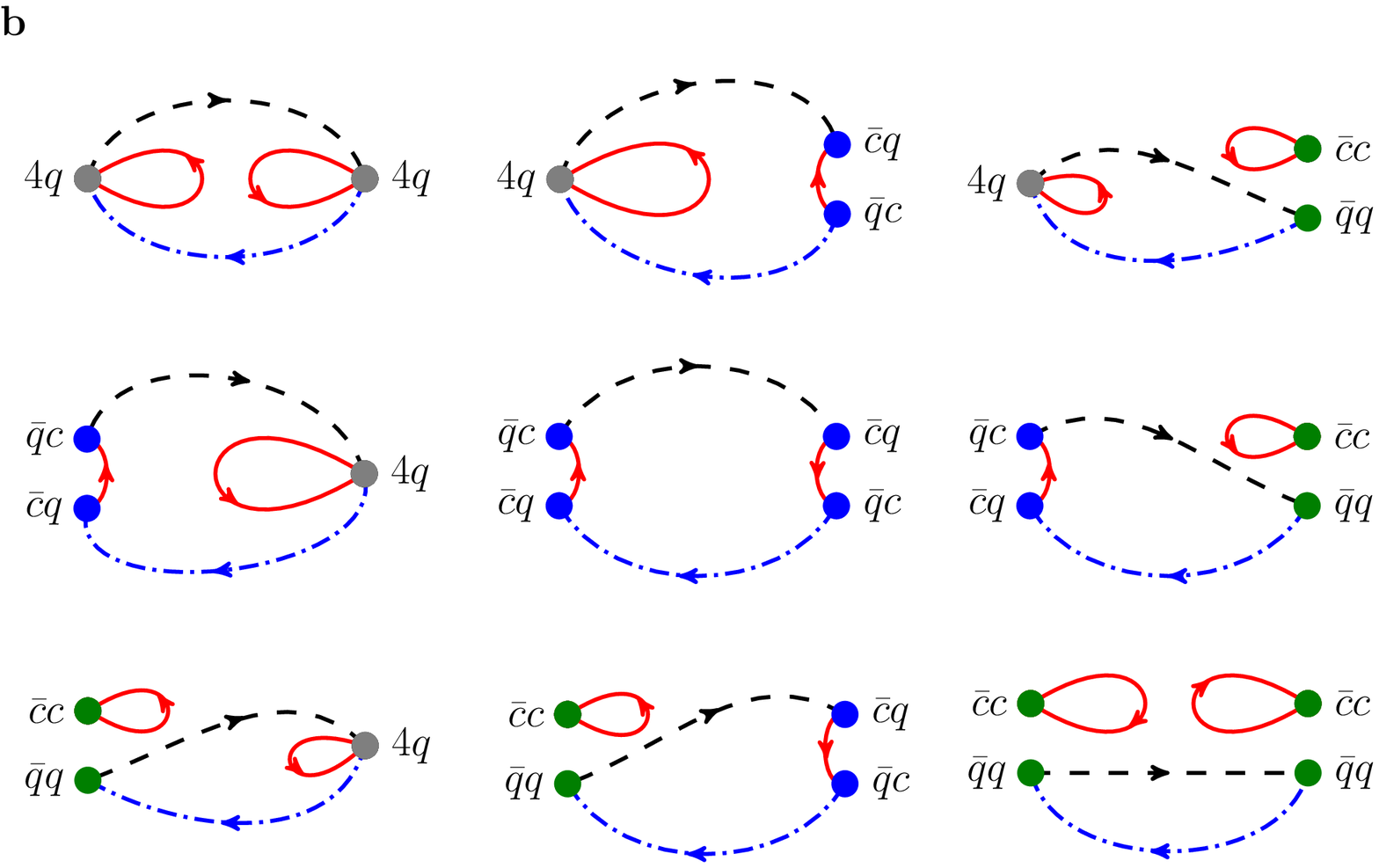}
    \caption{Diagrams for Wick contractions in the $I^G(J^{PC})=1^+(1^{+-})$ charmonium-like channel. Full red lines represent the charm quarks, dashed black lines the up quark and the dashed dotted blue lines the down quark. \textbf{a} Wick diagrams, where charm quarks propagate from source timeslice to sink timeslice. \textbf{b} Wick diagrams, where charm quarks propagate from source/sink timeslice to source/sink timeslice. As these diagrams couple to lighter states, they are not included in the analysis.}
    \label{fig:fig2}
\end{figure}

The Wick contractions obtained from these operators are shown in Fig. \ref{fig:fig2}. The contractions can be divided into two distinct types of diagrams, those where charm perambulators propagate from source to sink and those where they propagate from source to source or sink to sink. The latter couple strongly to light states with the same quantum numbers, which makes the extraction of any physical quantities at charmonium-like energies unfeasible and are thus not used in the calculation. Note that these diagrams are also omited in most lattice simulations of charmonium and other hidden charm states.\\

\vspace{-0.6cm}
\section{Results}
To obtain the spectrum from the correlator matrix we solve the Generalized EigenValue Problem (GEVP) \cite{Michael:1985ne,Luscher:1985dn,Luscher:1990ck,Blossier:2009kd} :
\begin{equation}
C(t) \vec{u}(t)=\lambda(t)C(t_0)\vec{u}(t).
\end{equation}
The energies are determined from the large time limit of the eigenvalues $\lambda(t) \propto e^{-E_{n}t}$ and the correlator matrix can be written as:
\begin{equation}
C_{ij}(t)=\sum_n Z_i^n Z_j^{*n} e^{-E_n t}.
\end{equation}
At first we looked at the basis ${\cal O}_{1-9,11,13-15,17,19-22}$\footnote{Results presented at the conference were founded on this basis and its sub-bases.}, where we identified a viable candidate for $Z_{c}$ at $4.16$ GeV. However after adding operators  ${\cal O}_{10,12,16,18}$, which correspond $\psi(2)\pi(-2)$, $\eta_c(1)\rho(-1)$, $D(2)D^*(-2)$, $D^*(1)D^*(-1)$ two-meson states, we found that our candidate was actually a linear combination of the two-meson states, which was missing in the smaller basis.

\begin{figure}[htb]
    \centering
    \includegraphics[width=0.7\textwidth]{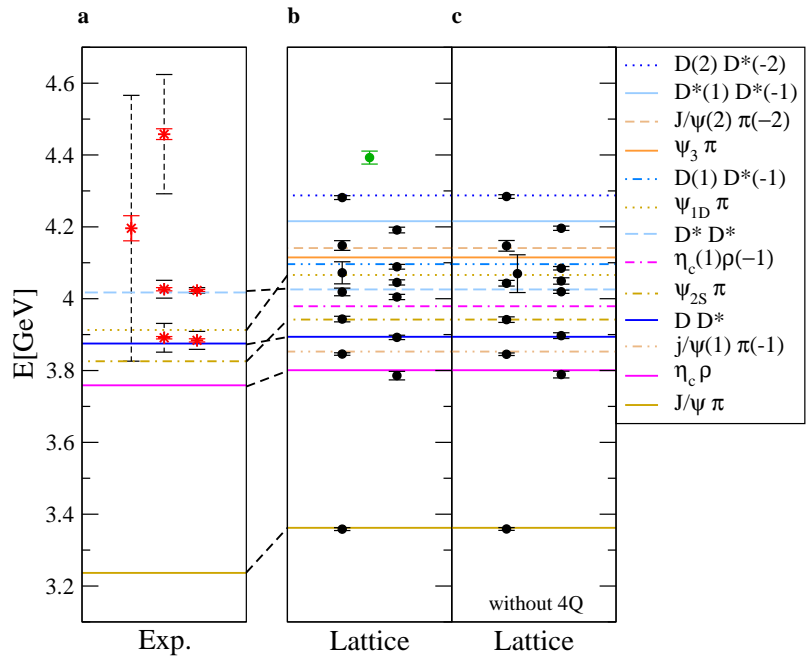}
    \caption{The spectrum for quantum numbers $I^G(J^{PC}) = 1^+(1^{+-})$ from the full basis. \textbf{a} Experimental status as in Table 1, dashed lines show the decay width of the state. \textbf{b} Results from full basis. Black dots correspond to two meson levels and the green dot is an additional state above $4.3$ GeV. It is likely another two-meson state not included in the basis. \textbf{c} Results from basis without diquark anti-diquark operators.}
    \label{fig:fig3}
\end{figure}

The fitted energies from the full basis are presented in Fig. \ref{fig:fig3}, where the experimental status is compared with ${\cal O}_{18-22}$ either included or excluded in the simulation \cite{Prelovsek:2014swa}. The black circles represent states identified as two-meson states; they lie near their expected noninteracting energy levels. The additional green circle shown in the middle plot of Fig. \ref{fig:fig3} could be a potential candidate. However as it is uncertain whether it is an additional state corresponding to a candidate, or just another two-meson level related to the tetraquark operators via Fierz transformations, we cannot reliably claim it to be a candidate. It lies at approximately $4.4$ GeV, which is outside of the region which we covered with two-meson operators. Further details on this analysis can be found in Ref. \cite{Prelovsek:2014swa}.


Note that an additional level is expected in elastic scattering, if the decay widths of the states are not too broad, and we have verified from experimental widths that the $Z_c$'s are narrow enough to give an additional state. Based on these results we conclude, that either there is no sizable $[\bar{c}\bar{q}]_{3}[cq]_{\bar{3}}$ component in the $Z_c$, that the pion mass is to high or that the assumption of the appearance of the additional level does not hold.\\
It is still unclear whether the $Z_c$ is a typical resonance, or if it only appears due to multi-channel phenomena. There is also another potential reason why $Z_c$ did not appear in our simulation; the $Z_c^+(3900)$ was found only in $e^+e^- \to Y(4260) \to (J/\psi \pi^+)\pi^-$ \cite{Ablikim:2013mio,Liu:2013dau,Xiao:2013iha}, but not in the decays $B_0^- \to (J/\psi \pi^+)K^-$ \cite{Chilikin:2014bkk},  $B_0^- \to (J/\psi \pi^+)\pi-$ \cite{Aaij:2014siy} or in $\gamma p \to (J/\psi \pi^+)n$ \cite{Adolph:2014hba}. This might indicate that the peek seen in $e^+e^- \to Y(4660) \to (J/\psi \pi^+)\pi^-$ might not be of dynamical origin.


\section{Acknowledgments}
We thank Anna Hasenfratz for providing the gauge configurations. S.P. thanks Changzheng Yuan and An\v ze Zupanc for discussion and D.M. is grateful for discussions with Jim Simone. We acknowledge the support by the Slovenian Research Agency ARRS project N1-0020 and by the Austrian Science Fund FWF project I1313-N27. Fermilab is operated by Fermi Research Alliance, LLC under Contract No. De-AC02-07CH11359 with the United States Department of Energy.


\end{document}